# MALWARE ANALYSIS USING MULTIPLE API SEQUENCE MINING CONTROL FLOW GRAPH

Anishka Singh[1]  Rohit Arora[1]  Himanshu Pareek[1]
*anishka.singh9@gmail.com*  *rohitarora089@gmail.com*  *himanshupareek@gmail.com*

[1]Centre for Development of Advanced Computing, Hyderabad, INDIA

## ABSTRACT

Malwares are becoming persistent by creating full- edged variants of the same or different family. Malwares belonging to same family share same characteristics in their functionality of spreading infections into the victim computer. These similar characteristics among malware families can be taken as a measure for creating a solution that can help in the detection of the malware belonging to particular family. In our approach we have taken the advantage of detecting these malware families by creating the database of these characteristics in the form of n-grams of API sequences. We use various similarity score methods and also extract multiple API sequences to analyze malware effectively.

## 1. INTRODUCTION

Malwares authors are applying various techniques like obfuscation, dead code insertion, creation of variants of one family for escaping the detection process. We focus on unpacked binaries for implementation of our approach. If a binary is packed with UPX (Ultimate Packer for eXececutables), then binary can be unpacked using UPX unpacker itself [15]. Otherwise if binary is packed with some other packer then that binary is executed in virtual environment and it's unpacking is done. After this step, open source BeaEngine library [16] is used to disassemble the executable's text section (figure 3). Disassembly is a process of decoding mnemonic codes into its equivalent assembly instructions. Algorithms applied for static dis-assembly falls in two categories.

1. Linear disassembly: In this disassembly starts at the first byte of .text section and instructions are decoded one after other.
2. Recursive disassembly: In this disassembly starts at the first byte of .text section and disassembly continues until a branch instruction is encountered. For a branch instruction control is transferred to the ad-dress in branch instruction and address of the next instruction is stored in stack. During disassembly if return instruction is encountered then control re-turns to the stored previous instruction's EIP. We use the BeaEngine Disassembler Library which utilizes recursive method of disassembly.

A Control Flow graph (CFG) of an executable is generated after disassembly. A Control Flow graph shows the flow of execution in the form of basic blocks (nodes) and edges. A basic block is a set of instructions without any branch or control transfer instruction in between. Edges are created between two blocks on the basis of branch or control transfer instruction. There are two types of control transfer instructions.

1. Conditional Control Flow Instruction This instruction transfers control to other instruction based on the condition being true or false. For example, JNE = jump if not equal.
2. Unconditional Control Flow Instruction This instruction includes instructions like JMP and call. JMP instruction transfers the control at a different instruction and control never returns to the same instruction. CALL instruction transfers the control to the instruction found in the called address and control returns to the calling instruction when a return instruction is encountered while executing instructions.

API database is created which includes Win32 soft-ware development kit's API as well as driver development kit's API. All the API's names are resolved in dis-assembly process by using Name or Hint array and Import Address Table present in PE structure [14]. CFG generation module generates Control Flow Graph from the Disassembly done by BeaEngine. All the Conditional and Unconditional jumps are taken care to de-ne a directed graph from the disassembly. The Directed Graph nodes contain the Win32 API call which is the execution sequence of the executable disassembled. Depth First Search algorithm is applied for traversal of all the nodes through which Multiple API call sequences are retrieved. We prefer depth first traversal over breadth first traversal because former will be able to build a path while going down the depth. How-ever BFS will traverse all nodes at one level before going to next level and hence limit its capability to build the path automatically during traversal. To analyze features of API sequences, we used N-gram technique and calculated 2-gram, 3-gram and 4-gram sequences from the obtained API sequences. Using a large training dataset, Malware N-Gram database and Benign N-Gram database are generated for all gram factors. We have used coefficients like Dice coefficient, Cosine coefficients and Tversky Index for calculating the similarity between the N-Gram databases generated from the training database and the N-Grams obtained from the file under analysis. These similarity measures help in deciding whether the executable under analysis is malicious or benign. Formulae for dice, Tversky and cosine coefficients are shown in equations 1, 2, 3 respectively. X and Y are two sets in which we have to determine the similarity.

$$D = \frac{2\,|X \cap Y|}{|X|+|Y|} \quad \text{(Eq.1)}$$

$$T = \frac{|X \cap Y|}{|X| \cap |Y| + \alpha|X-Y| + \beta|Y-X|} \quad \text{(Eq. 2)}$$

Where α and β are min and max values of |X-Y| and |Y-X|.

$$C = \frac{2\,|X \cap Y|}{\sqrt{X+|X \cap Y|*(Y+|X \cap Y|)}} \quad \text{(Eq.3)}$$

If a file is found malicious then there are many classes under which that particular malicious file be-longs too. Characteristics or traits present inside the malicious file always help in depicting the class of that particular malicious file. We have classified the malware into different malware classes i.e. Trojan, Back-door, Viruses, Adware, and Worms. So, the database of each malware class is a subset to Malware N-Gram database. If the file is found malicious then we can also tell the user the name of malware class in which it belongs.

## 1.1 Problems in Evaluation of Packed Executable Detectors

Unfortunately there is no simple way to deterministically find out whether a binary is packed or not [19].

We have to take multiple steps in evaluating packed binary detectors [20]. Many papers evaluate their packed executable detection approaches and post a true positive rate and false positive rate. But accuracy of these rates can be questioned as accuracy of data sets used is not validated. When data set is built by authors themselves by packing non packed executables with widely known packers, accuracy is trusted but approach has not been put to rigorous testing on real malware picked from internet [21][22]. When a data set is built by obtaining real malware samples, then we basically don't know which binaries are packed and which are not packed [17]. In this case accuracy cannot be trusted. We designed and implemented a packed executables detection approach based on dynamic analysis [23][24][25]. Our tool utilizes PIN tool for instrumentation [26]. When a packed binary unpacks itself then it should insert some decrypted code into program's address space and should execute the same. By utilizing PIN we have written a tool to trace down all the written address of a running program and if the program tries to execute from the same written address then we flag it as packed[27][28]. When the results obtained by using this approach were cross validated with the results obtained from Signature based approaches (e.g. PEiD)[29], we found many instances which were packed but could not be detected by dynamic analysis based approach. The problem is when we don't know exactly which binaries are packed, we cannot trust either static analysis based detection [30][31] or dynamic analysis based one. Hence we must first start with something trusted and extend the chain of trust. Michael St. Neitzel emphasizes the problems in packed executable detection testing [32] For example when a binary is flagged as non-packed even by dynamic analysis approach it may be due to binary did not execute in its full capability because of various dependency and checks. We propose a systematic methodology for evaluating packed executable detectors. The approach is shown in Figure 1. Here we list down involved steps. In first step we collect and divide executables on the basis of following two methods. a) Packed Binaries which are correctly detected by signature based approaches [29] and b) Packed Binaries which are not detected by signature based approaches but either packed program or packer itself is available. So, multiple binaries of this type will be collected. In second step, we develop a dynamic analysis based tool for detecting packed binaries. Now the set of binaries which is built in first step is fed to this tool to observe the capability of dynamic analysis based detection and these results can be crucial base for benchmarking output from next step. In third step, take a large dataset of binaries on which packed binary detectors has to be tested. These binaries now can be fed to dynamic analysis based tool for classifying packed and non-packed binaries. Moreover, detection must be applied for binaries which refused to run in their full capability. Hence binaries which are VM-aware and Debugging aware can be weeded out. In last step, static analysis based detection tool can run on packed binaries which are declared as packed by our previous step. This Approach helps to build a validated data set resulting accurate detection rate.

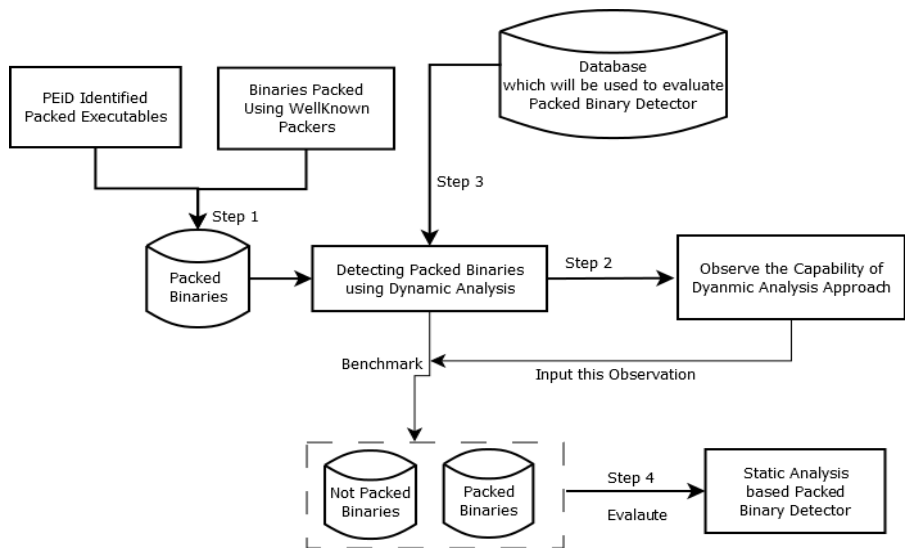

Figure 1: Proposed Approach for Evaluation of Packed Executable Detectors

## 2. RELATED WORK

Sung et al. [1] has introduced various obfuscation techniques that can be applied on malware samples and then malware signatures based on the API sequences are generated from these obfuscated binaries. New suspicious binaries are scanned with this newly created malware signature database. With the help of similarity measure like Euclidian distance and similarity functions like Cosine measure, extended Jaccard measure, and the Pearson correlation measure these suspicious binaries are classified as malicious or benign. Li et al. [2] has developed a PE parser for extracting API sequence and classified the PE as benign or malicious by using association rules. Association rules are generated based on OOA Fast FP-Growth algorithm which is based on support and confidence functions. Similarly Ban et al. [3] has proposed inserting hooks into the running program and extracting the critical APIs that are categorized on the basis of their functionality. Eskandari and Hashemi [4] have also proposed a feature selection algorithm by assuming called APIs on the CFG. Similarly Zhao [5] has designed a virus detection model based on feature selection in Control Flow Graph and generating classifiers according to specific machine learning algorithms. Yang [6] has extended the support for detection using structural features generated from CFG by identifying the packed binaries with the help of entropy. These packed binaries are unpacked in dynamic environment. Similarly Canzanese et al. [7] has measured the kernel APIs call and sequence of API call made by running the binaries in virtual environment and classifying them as malicious or benign by applying random forest classifier. Bonfante et al. [8] has generated malware signatures by constructing the control ow graph based on opcode like JMP, JCC, CALL and RET instructions forming the sequence in CFG. Mithun et al. [9] has demonstrated the correct and effective API usage by mining the API partial order. Watters et al. [10] proposed behavior analysis using different behavior groups of API call features extracted from IDA Pro tool. Similar approach is followed by Veeramani et al. [11] by doing unpacking of known packed binaries with respective packers and extracting API sequence of unpacked binaries with the help of IDA Pro. Super-vised learning based SVM classifier has classified binaries into benign or malicious. Wasaki et al. [12] has created their own recursive disassembler and constructed control ow graph to extract API sequence and used dice coefficient to calculate similarity between malware samples. Parvez, et al. [13] has used ether virtualization for unpacking the packed binaries and applied various machine learning algorithms on API n-gram extracted from the disassembly of the unpacked binary.

## 3. OUR APPROACH

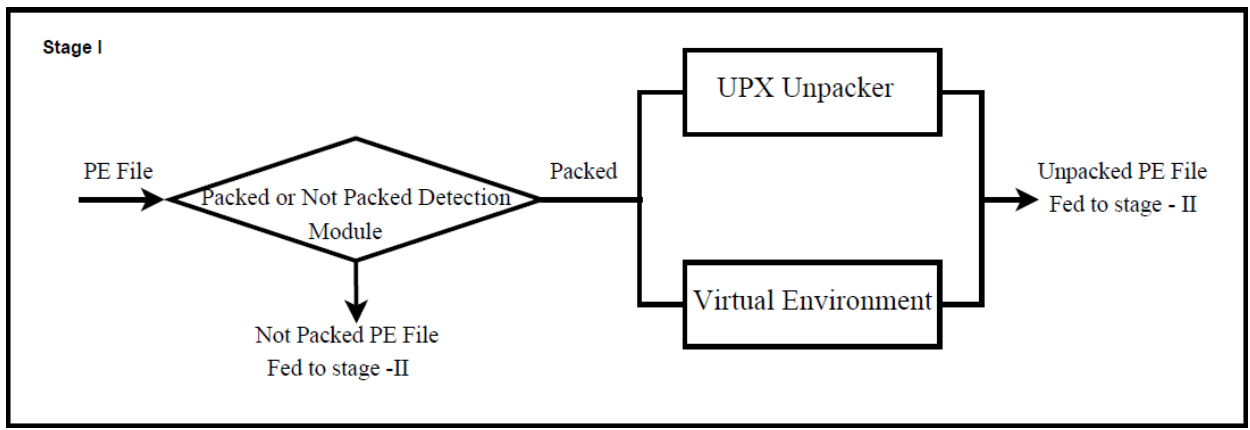

Figure 2: Fitting Packed Binary Detection in overall approach

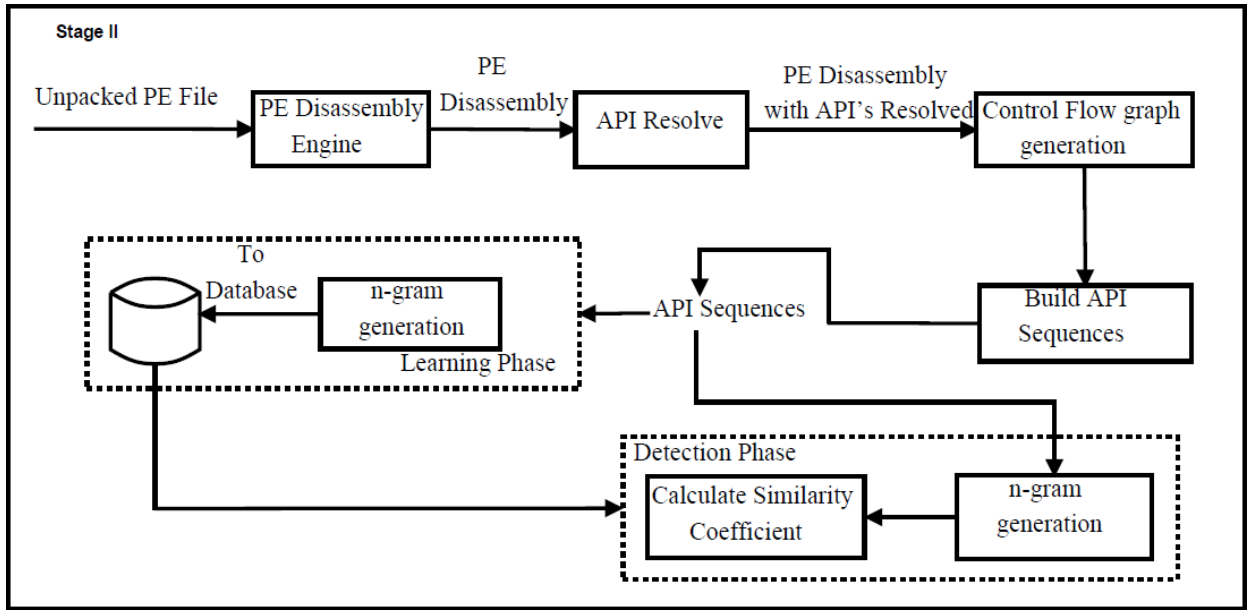

Figure 3: Detection Approach

### 3.1 Resolving API in Disassembly

After disassembly API calls made in the disassembly are resolved by using Name/Hint array and Import Address Table present in PE structure. For e.g. CALL DWORD PTR [145234]. In above instruction call address i.e. 145234 comprises of base address for the PE i.e. 0x140000 and the thunk value 5234. This thunk value is matched against the thunk value present in Import Address Table read from Portable Executable. If a match is found then the Function Name present in Import Address Table is assigned to this Instruction. API database is used which contains API's from Win32 software development kit as well as driver development kit for further restricting the Function name resolved in the disassembly.

### 3.2 Control Flow Graph generation

In control ow graph generation for portable executables conditional and unconditional instructions are used for creating edges between the nodes in CFG. When a program is running and a conditional instruction is encountered then it follows single path depending on the condition being true or false. But during static analysis dynamic behavior of the program cannot be deter-mined. Hence, while generating CFG whenever any conditional instruction is encountered then both paths are taken for the creation of nodes. Address of entry point found in the text section of a PE is taken as the starting point for the generation of CFG. From the entry point all instructions are taken one by one and placed inside node until a branch instruction is encountered.

Conditional instruction
If conditional branch instruction is encountered then either that branch will be taken or the program will follow the normal ow by executing the next EIP. If branch is taken then address present in the instruction is matched with the disassembled instruction's EIP and different node is formed. Next EIP of the branch instruction is stored in the stack. Whenever control re-turns from the branch instruction then EIP stored in the stack is popped for creation of nodes.

Unconditional instruction
If unconditional JMP instruction is encountered, then the address present in the jmp instruction is matched with disassembled instruction's EIP and the new node is formed. If address is not matched then CFG creation stops as for JMP instructions execution never return to next EIP. But if the encountered instruction is call instruction then the called address is matched with dis-assembled instruction's EIP and a new node is formed. The control is again transferred to the next EIP of the calling instruction whenever a return instruction is encountered. In this way all the nodes along with links are created in the CFG. Nodes having conditional branch instruction like jne or unconditional branch instruction like call will have two adjacent nodes linked with them. For nodes with unconditional branch instruction like JMP will have only one node linked with them as for them there is only one path. There also exist instructions such as CALL RAX, JMP EAX etc. where a register value is used to provide jump address to instruction. This approach will require some sort of emulation or symbolic execution to calculate dynamic value in the operand. These kinds of cases are not currently handled in our work.

### 3.3 Extracting Multiple-API sequences using Depth-First traversal

Depth First Traversal is used for traversing all the linked nodes till return instruction is found and then back-tracking is done. For backtracking in CFG the address stored in the stack is popped and traversal is started from that address. We are using vector for storing the entire path traversed before backtracking. All the nodes in CFG will have adjacent nodes which are marked as Not Visited. So, one of the adjacent nodes is taken during traversal and it is stored in vector and marked as visited. The other adjacent node's EIP is stored in the stack. Whenever a return instruction is encountered then the previously traversed path instructions which are stored in vector makes one path and traversal again starts from the previously stored adjacent node EIP. This procedure is followed until all the paths present in the CFG are traversed and stored in the vector. All the branch instruction traversed in the CFG which forms one of the paths is stored inside a vector. These paths also include the API calls made while creating the CFG and forms a sequence of API calls made in one of the many paths.

### 3.4 Feature extraction

N-gram sequence algorithm
After extraction of multiple API sequences from the portable executable, n-gram technique is implemented for analyzing these API sequences. A large set of known malware is taken and divided into five major classes (Virus, Trojan, Backdoor, Adware and Worm). Later on n-grams are extracted from the API sequences obtained from these categories of malware dataset. Thus n-gram construction leads to five different malware n-gram databases. And one database for benign dataset is also constructed by taking a large set of benign les from windows system directory.

Many malware detection approaches based on API sequence has extracted the API's from the executable and constructed n-grams from them. Later on they have applied some machine learning algorithm like random forest, Naive Bayes, J48 on them and provided the out-put as benign and malicious. In these kinds of approaches each API extracted from the API sequence is assigned some unique identifier as shown in table 1. For example, suppose API sequence extracted from a file is 1, 2, 3, 4, 5, 6, 7, 8, 9, 10, 11, 12. For malicious file 3-grams is represented as shown below:

{1, 2, 3, Malicious}
{2, 3, 4, Malicious}
{3, 4, 5, Malicious}
{4, 5, 6, Malicious}
{5, 6, 7, Malicious}
{6, 7, 8, Malicious}
{7, 8, 9, Malicious}
{10, 11, 12, Malicious}

At other places in the paper we show only a fraction of these n-grams (wherever required) to be concise.

Table 1: Listing API Name and associated Unique Id

| API Name | Unique Id | API Name | Unique Id |
|---|---|---|---|
| ReadFile | 1 | GetProcAddress | 7 |
| WriteFile | 2 | VirtualAlloc | 8 |
| CloseFile | 3 | VirtualAllocEx | 9 |
| OpenFile | 4 | FindFirstFile | 10 |
| CreateFile | 5 | FindNextFile | 11 |
| CreateProcess | 6 | LoadLibrary | 12 |

Similarly, 3-grams extracted from single benign file is represented as shown below:

{1, 2, 3, Benign}
{4, 5, 6, Benign}
{7, 8, 9, Benign}
{10, 11, 12, Benign}

Later on training is done on these created 3-grams by Weka tool [18]. This type of dataset doesn't provide much accurate detection. As we can see from the table 2 we have taken 51 samples from Trojan-Spy.Win32.Zbot class out of that 26 samples were used for training and 25 samples were used for testing. In 25 testing samples 6097 3-grams were extracted in which 2811 3-grams were detected and 3286 3-grams were not detected by using random forest algorithm in Weka tool. Weka by default uses 10-fold cross validation. We are presenting a set based approach in which API sequence extracted from a file is 1, 2, 3, 4, 5, 6, 7, 8, 9, 10, 11, 12 then for a malicious file 3-grams extracted are represented as shown below:

{(1,2,3),(4,5,6),(7,8,9),(10,11,12), Malicious}

Similarly, 3-grams extracted from a benign file are represented as shown below.

{(1,2,3),(4,5,6),(7,8,9),(10,11,12), Benign}

The results obtained by using this approach are rep-resented in table 3. We have taken 20 samples from Win32-Genome-Trojan malware class out of which 10 samples are used for training the model and 10 samples are used for testing. By using random forest algorithm all the 10 samples used in testing are correctly identified by our set based approach for 3-grams.

Table 2: Table showing Results obtained from gram based approach:

| S. NO | Malware Family | Samples Count | Training Sample Count | Test Sample Count (n-gram extracted) | Random Forest Algorithm | |
|---|---|---|---|---|---|---|
| | | | | | D | ND |
| 1 | Trojan-Spy.Win32.Zbot | 51 | 26 | 25 (6097) | 2811 | 3286 |
| 2 | Win32-Genome-Trojan | 20 | 10 | 10 (13120) | 13020 | 100 |
| 3 | Win32-Inject-Trojan | 10 | 4 | 6 (684) | 144 | 540 |
| 4 | Win32-Palevo-p2p-Worm | 10 | 5 | 5 (938) | 461 | 477 |
| 5 | Win32-ZAccess-Backdoor | 22 | 11 | 11 (935) | 67 | 868 |
| 6 | Win32-Agent-Trojan-Dropper | 6 | 3 | 3 (331) | 65 | 266 |

Note  1 *D = Detected n-gram sequence,
      *ND = Not Detected n-gram sequence

Similarity Coefficient

Different Similarity coefficients are used to calculate the similarity between file under analysis and malware, benign n-gram databases. Similarity is calculated on the basis of gram sequences extracted from the file under analysis. If the similarity coefficient with malware API n-grams database is higher, it means sample file contains features similar to malware and the coefficient value tells to which category of malware class it belongs too.

Table 3: Table showing Results obtained from set based approach

| S. NO | Malware Family | Samples Count | Training Sample Count | Test Sample Count (n-gram extracted) | Random Forest Algorithm | |
|---|---|---|---|---|---|---|
| | | | | | D | ND |
| 1 | Trojan-Spy.Win32.Zbot | 51 | 26 | 25 | 17 | 8 |
| 2 | Win32-Genome-Trojan | 20 | 10 | 10 | 10 | 0 |
| 3 | Win32-Inject-Trojan | 10 | 4 | 6 | 1 | 5 |
| 4 | Win32-Palevo-p2p-Worm | 10 | 5 | 5 | 1 | 4 |
| 5 | Win32-ZAccess-Backdoor | 22 | 11 | 11 | 7 | 4 |
| 6 | Win32-Agent-Trojan-Dropper | 6 | 3 | 3 | 2 | 1 |

## 4 EXPERIMENTAL RESULTS

A large training dataset of 15000 malwares and 4000 benign PE les is collected. With the help of this dataset, we have built 2-gram, 3-gram and 4-gram databases of API sequences as our training database for five categories of malware classes and benign files. We have identified the malware classes by uploading those samples on virus total. These training n-gram databases are further used for testing and calculating the similarity between extracted API sequence n-gram from file under analysis and the database built. These similarity coefficients tell the file under analysis is more similar to malware indicating malware characteristics or similar to benign indicating benign characteristics.

For testing we have taken 2340 samples for 2-grams, 4672 sample for 3-grams and 2416 samples for 4-grams. By taking three similarity coefficient i.e. Dice, Tversky and cosine we have done comparison of the performance of these coefficient in detection of the malware. Results obtained from the given testing dataset are displayed in figure 4 and 5. Detection rate of 94.78% and false positive rate of 33.51% has been achieved by using 3-grams and dice coefficient. After this we have applied a cumulative approach for 3-grams with Dice coefficient.

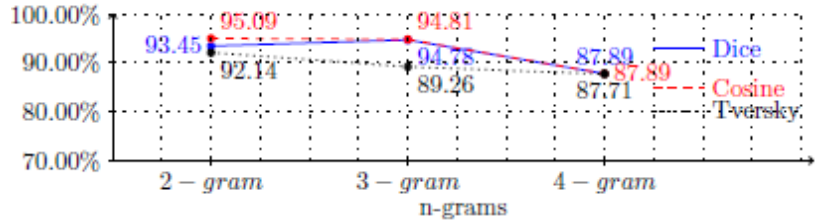

Figure 4: TPR for 2-grams, 3-grams and 4-grams

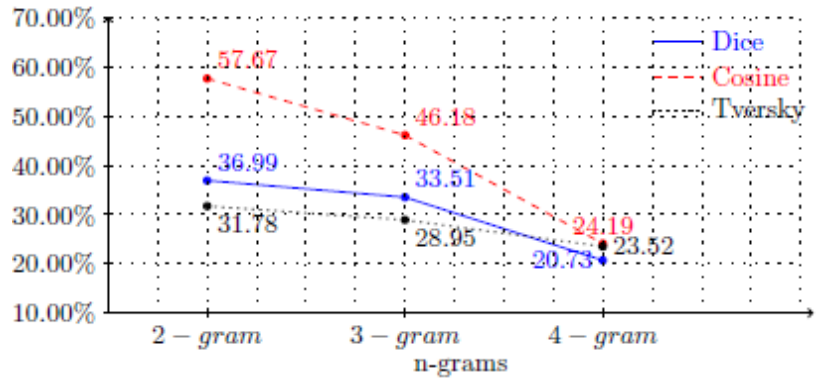

Figure 5: FPR for 2-grams, 3-grams and 4-grams

This approach comprises of four phases. In phase-0 we have taken training set of 12443 malwares and 11561 benign and constructed the n-gram database. In phase-1 we have taken new 3500 malware samples and 3094 benign samples and per-formed the testing of the constructed database. By this detection rate of 92.00% has been achieved. Later on we have done cumulative learning by adding these samples in the previous training dataset thereby increasing the learning set to 15943 malicious samples and 14655 benign samples. In phase-2 new 3500 malware samples and 2767 benign samples are taken and testing is done on the database obtained from the phase-1 cumulative learning. By this detection rate of 94.04% is achieved. Again these samples are added into the database constructed from phase-1 learning thereby making the over-all database comprising of 19443 malware samples and 17422 benign samples. To build these training sets, it required at least 1m per malware. While testing a given binary requires it requires around 3m per malware.

In phase-3 similar approach is followed by taking new 3278 malware samples and 1689 benign samples and testing is done on the database obtained from phase-2 cumulative learning. Detection rate of 95.06% is obtained by following this approach. In this phase also these samples are added into the database and now the database comprises of 22721 malicious samples and 19111 benign samples. Figure 6 represents the cumulative approach for learning the database in respective phase-0, phase-1, phase-2and phase-3. At each phase test, we obtain malware set which is not detected and benign set which is wrongly detected. We add these misidentified les and prepare updated database for next phase. Our cumulative testing approach asserts the fact that the technique improves with more training data.

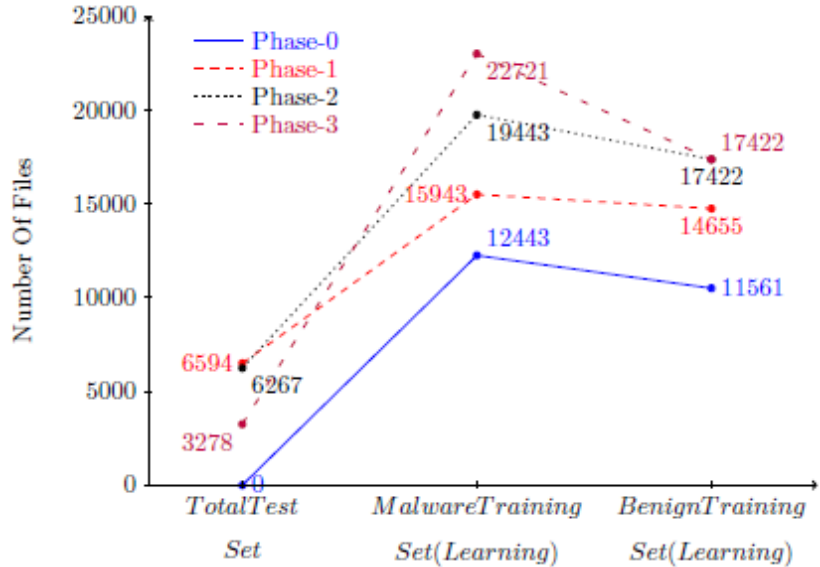

Figure 6: Malware and benign dataset for cumulative testing and training

Detection rate achieved in various phases in depicted in figure 7 and false positive rate achieved in various phases in depicted in figure 8. False positive rate can further be reduced by using more benign binaries to build training set. Every malware belongs to one malware family and share some traits of that family. For testing these traits similarity we have taken six malware families and per-formed phase wise learning by taking 1, 2 and 3 sample files from each of the malware families. The detection rate obtained by using this approach is depicted in table 4.

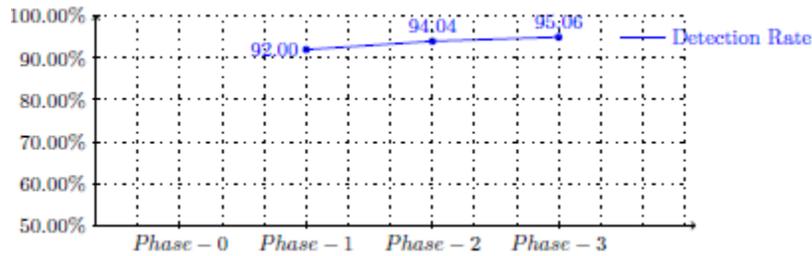

Figure 7: Detection Rate in Phase I, II, III

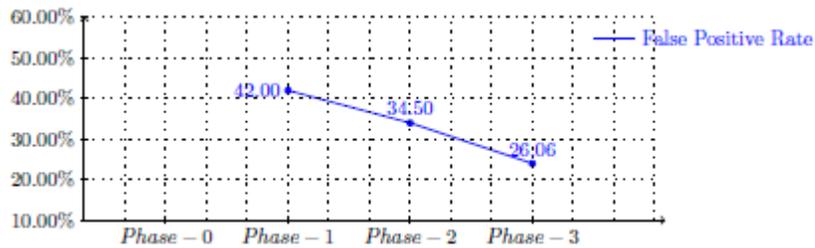

Figure 8: False Positive Rate in Phase I, II, III

We can see from table 4 that when we take 1 sample from the respective malware family and test our samples for the similarity with that particular family then detection rate is bit lower. In comparison to that when sample count is increased to 3 file then detection is approx. 100% in three families. This gives the idea that malware belonging to one malware family shares approximate similar API sequences.

Table 4: Detection rate based on 1, 2, 3 Files used as training dataset in Phase I Phase II and Phase III

| S. NO | Malware Family | Samples Count | Phase I 1 Files | Phase II 2 Files | Phase III 3 Files |
|---|---|---|---|---|---|
| 1 | Trojan-Spy.Win32.Zbot | 51 | 21.56 % | 52.94% | 68.62% |
| 2 | Win32-Genome-Trojan | 20 | 100% | 100% | 100% |
| 3 | Win32-Inject-Trojan | 10 | 40.00% | 90.00% | 100 % |
| 4 | Win32-Palevo-p2p-Worm | 10 | 30.00% | 60.00% | 70.00% |
| 5 | Win32-ZAccess-Backdoor | 22 | 40.90% | 59.09% | 72.72% |
| 6 | Win32-Agent-Trojan-Dropper | 6 | 50.00% | 83.34% | 100% |

# 5 CONCLUSIONS

In this paper we have presented the accuracy of the set based n-gram detection technique using multiple API's sequence. In our work we came out with a cumulative approach for testing and training our databases and hence improving the detection rate. We have also presented an approach that detects malware and its variant belonging to different classes. We see that our approach is very useful in detecting variants of one malware family. We also observe that some mathematical research to determine more efficient similarities can be taken up in future.